\documentclass[aps,preprint,showpacs]{revtex4}
\usepackage{graphicx}

\begin{document}

\title{Elastic properties of small-world spring networks}

\author{A. Ramezanpour}
\email{ramezanpour@iasbs.ac.ir}

 \affiliation{Institute for Advanced Studies in Basic Sciences,
Zanjan 45195-1159, Iran}

\author{S. M.
Vaez Allaei} \email{smvaez@iasbs.ac.ir}

\affiliation{Institute for Advanced Studies in Basic Sciences,
Zanjan 45195-1159, Iran}

\date{\today}

\begin{abstract}
We construct small-world spring networks based on a one
dimensional chain and study its static and quasistatic behavior
with respect to external forces. Regular bonds and shortcuts are
assigned linear springs of constant $k$ and $k'$, respectively. In
our models, shortcuts can only stand extensions less than
$\delta_c$ beyond which they are removed from the network. First
we consider the simple cases of a hierarchical small-world network
and a complete network. In the main part of this paper we study
random small-world networks (RSWN) in which each pair of nodes is
connected by a shortcut with probability $p$. We obtain a scaling
relation for the effective stiffness of RSWN when $k=k'$. In this
case the extension distribution of shortcuts is scale free with
the exponent $-2$. There is a strong positive correlation between
the extension of shortcuts and their betweenness.  We find that
the chemical end-to-end distance (CEED) could change either
abruptly or continuously with respect to the external force. In
the former case, the critical force is determined by the average
number of shortcuts emanating from a node. In the latter case, the
distribution of changes in CEED obeys power laws of the exponent
$-\alpha$ with $\alpha \le 3/2$.
\end{abstract}

 \maketitle

\section{Introduction}\label{1}
Recent studies indicate that real networks have a complex
structure and function \cite{ab,n1,ws,asbs,bw}. The small-world
network (SWN) introduced by Watts and Strogatz \cite{ws} captures
some basic ingredients of real networks. There are lots of studies
dealing with static properties of SWNs, see \cite{n1} and
references therein. However, we still only know a little bit about
dynamical features of small-world networks
\cite{ah,slr,cvv,aks,mhmk,mak}. To best of our knowledge there is
no study on the elastic properties of SWNs. Certainly the most
physical application of small-world networks is in the context of
macromolecules and polymers \cite{jsb1,jsb2,jb,vdpk,kk}. The
interaction pattern of a polymer can be represented by a SWN.  In
this way, the monomers are mapped into nodes of a network and the
interaction between two spatially close monomers is shown by a
bond between the corresponding nodes. Note that here there are two
kinds of bonds: regular and long range. The long range bonds that
are usually weaker than the regular ones, could stand much smaller
stresses and would be broken more easily. If we model the
interactions with linear springs, we will obtain a small-world
spring network.  Of course, we should take more care on modeling
linear polymers with SWNs \cite{sc}. Nevertheless, we expect that
the study of small-world networks (as a toy model of polymers when
thermal activities are absent) provides
insights about more complicated behaviors of real polymers.\\
Suppose that we have a chain of elastic fibers and add some other
fibers to randomly selected pairs of nodes. Then it is
interesting to know , for example, the effective stiffness of this
object and some other quantities which are of interest in
fiber-bundle models \cite{s,hh}.\\
In this paper we are going to study the behavior of small-world
spring networks when external force $F$ is exerted on the end
nodes. The network response contains some information about its
internal structure. For a given $F$,  we obtain the network
stiffness and the extension distribution of long range bonds
(shortcuts). Then we increase $F$ quasistatically and define a
cutoff length for shortcuts beyond which we remove them from the
network. During this quasistatic process the chemical end-to-end
distance (CEED), defined as the number of bonds in the shortest
path connecting node $1$ to $N$, could have nontrivial behavior
with $F$. Moreover by increasing $F$ one encounters a number of
avalanches in which one has a change in the number of shortcuts
and also in CEED. Certainly, distinct structures could lead to
different behaviors, for instance in the distribution of changes
in CEED. These
differences provide us a useful measure to classify various networks.\\
Here we will show that the extension distribution is scale free in
the small-world regime in contrast to the three-modal distribution
of complete-network regime. We obtain a positive correlation
between the betweenness \cite{n1} of a shortcut and its extension.
There is also a threshold value $F_c$ which leads to an abrupt
change in CEED. The scaling of $F_c$ with the size of the network
depends strongly on the network structure. Depending on the
elastic properties of
springs we could also have a continuous change in CEED.\\

The structure of this paper is as follows. After giving some
general definitions, we introduce and study a hierarchical
small-world network in section \ref{3}. In section \ref{4} we
study complete networks as another simple case. In section \ref{5}
we present the results of our numerical simulations and scaling
arguments for random small-world networks . Finally we give the
conclusion remarks.
\section{General definitions}\label{2}
We take a one dimensional chain with $N$ nodes, numbered $1$ to
$N$. This chain has $N-1$ regular bonds, with spring constant $k$.
The network structure is completed by adding some shortcuts of
spring constant $k'$. In the following we will consider the case
$0\leq k'\leq k$ which is more reasonable in physical models. The
effective elastic constant of the network is denoted by $K$. We
exert force $F$ on the end nodes of a chain and obtain the
extension distribution of shortcuts, $P(\Delta_x)$, where
$k'\Delta_x$ is the force acting on a shortcut. On increasing $F$
quasistatically, we define a cutoff length $\delta_c$ for
shortcuts. It means that a shortcut is teared if its extension exceeds $\delta_c$.
 Here we will take $\delta_c=1$ (the specific value of
$\delta_c$ does not affect the qualitative behavior of our
results). For a given network, we may have a number of avalanches in some particular forces.
 An avalanche starts by the tearing of a shortcut and ends when all the shortcuts
have an extension smaller than $\delta_c$. By increasing $F$, a
number of avalanches may occur which we label by index $a$. In
each avalanche, one can measure, for example, the force value
$F_a$, the chemical end-to-end distance $R_a$ and the effective
spring constant of the network $K_a$. The CEED, $R$, is defined as
the number of bonds in the shortest path connecting node $1$ to
$N$. Just after the $a-$th avalanche, the physical end-to-end
distance, $X_a$, is given by $F_a/K_a$. We also define the change
in a quantity such as $R$ by $\Delta R_a:=R_{a+1}-R_a$.
\begin{figure}
\includegraphics[width=8cm]{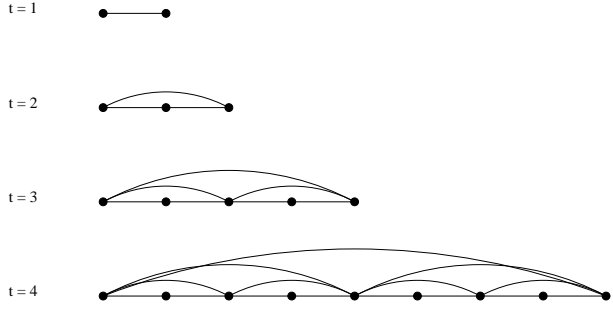}
\caption{ Constructing a hierarchical small-world network.
}\label{f1}
\end{figure}
\section{A hierarchical small-world network}\label{3}
Let us first study a simple hierarchical small-world network where
its behavior can be treated exactly. The construction of the
network is depicted in Fig. \ref{f1}. In the first step, $t=1$, we
consider two nodes connected with a regular bond. In the next
step, $t=2$, we make a copy of the previous step and merge them.
We also add an additional shortcut between the first and last
nodes of the new network. If we repeat this procedure for $t$
steps we obtain a hierarchical SWN of size $N(t)=2^{t-1}+1$ with
$M(t)=2^{t-1}-1$ shortcuts. One can obtain the following relation
for the effective stiffness of the network
\begin{equation}
K(t)=k'+\frac{K(t-1)}{2}=\frac{k}{N(t)-1}+2k'(1-\frac{1}{N(t)-1}).
\end{equation}
If we pull the end nodes of the network with force $F$ we find
that there are $2^{l}$ shortcuts of the extension
$\Delta_x=F/(K(t)2^l)$. Here $l \in[0, t-2]$ is an integer that
labels the shortcuts according to the step they have been added to
the network. This in turn results in the following extension
distribution
\begin{equation}
P(\Delta_x)\sim (\Delta_x)^{-2}.
\end{equation}
Now we start to increase $F$ from zero. In very small $F$,
$R_0=1$. When the force reaches to $F_1=K(t)\delta_c$, the spring
connecting node $1$ to $N(t)$ extends by $\delta_c$ and tears.
After this event, $R_1=2$ and the number of shortcuts decreases to
$M_1=M(t)-1$. Moreover the effective network stiffness is given by
$K_1=K(t-1)/2$. It is easy to see that a shortcut connecting two
far nodes has a larger extension than one connecting two near
nodes. Thus we can summarize the behavior of interesting
quantities versus $a$
\begin{eqnarray}\label{hr}
F_a=K(t-a+1),\\ \nonumber R_a=2^a,\\ \nonumber  M_a=M(t)+1-2^a,
\\ \nonumber K_a=\frac{K(t-a)}{2^a}.
\end{eqnarray}
The physical end-to-end distance then reads
\begin{equation}
X_a=\frac{F_a}{K_a}=R_a\frac{K(t-a+1)}{K(t-a)},
\end{equation}
which for $t-a\gg 1$ is approximately equal to $R_a$. Thus in this
limit the physical end-to-end distance is well given by CEED. From
the above relations we obtain $F_a=K_{a-1}R_{a-1}$. Consider the
bonds of the shortest path connecting node $1$ to $N$. $F_a$ is
the force at which the extensions of these bonds become
$\delta_c$. Note that, however, the relation between
force and CEED is linear
\begin{equation}
F_a=\frac{k R_a}{2^t}+2k'(1-\frac{R_a}{2^t}).
\end{equation}
From Eqs. (\ref{hr}) one easily finds that
\begin{equation}\label{fa}
\Delta R_a=-\Delta M_a=2^a,\hskip 1cm \Delta
F_a=\frac{k}{2^{t-a}}(1-\frac{2k'}{k}).
\end{equation}
We see that for $k'< k/2$, $F_a$ increases versus $a$. Thus in
this situation we have a continues change in $R$. In the continuum
approximation we obtain
\begin{equation}
P(\Delta R)\sim (\Delta R)^{-1},
\end{equation}
On the other hand,  when $k'>k/2$, from Eqs. (\ref{hr}) and
(\ref{fa}) we have $F_1=K(t)$ and $F_1>F_2>F_3\ldots$. It means
that after tearing the first shortcut, we reach a new
configuration in which the extension of most extended shortcut(s)
is(are) greater than $\delta_c$. This process continues until all
shortcuts tear. Note that during this event which occurs at
$F=F_1$, CEED changes from $1$ to $N-1$. Thus we have an abrupt
change of CEED at $F_c=K(t)$.
\section{Complete networks}\label{4}
In a complete network, each node is connected to all the other nodes.
Thus the number of shortcuts is $M_0=(N-1)(N-2)/2$. We start with
the simple case of $k'=k$ with the effective network stiffness of
$K_0=kN/2$. Moreover, $P(\Delta_x)$ consists of three
delta peaks corresponding to three different kinds of shortcuts in
the network: The first kind is a single shortcut between the end
nodes and has the largest extension. The second kind of shortcuts
which have less extensions, are those that connect the end nodes
to the inner nodes. And the remaining shortcuts, with zero extensions,
are the shortcuts connecting the inner nodes to each other.\\
In the absence of external force, $R_0=1$. For a given $F$ we have
the largest extension in the bond connecting node $1$ to $N$.
Hence by increasing $F$ the first avalanche occurs at $F_1=kN/2$.
At this step we have $M_1=M_0-1$, $R_1=2$ and $K_1=k(N-2)/2$. A
simple calculation shows that for all springs connected to nodes
$1$ and $N$, we have $\Delta_x=F/(k(N-2))$. So by increasing $F$
we reach $F_2=k(N-2)$ in which $2(N-3)$ shortcuts are teared. In
this step, we will have a complete network consisted of all nodes
between $2$ and $N-1$. But we know already that the spring
connecting node $2$ to $N-1$ can only stand forces lower than
$k(N-2)/2$ which is much smaller than $F_2$. Thus, in the second
avalanche, all shortcuts will be removed. So $F_c=k(N-2)$ is the
threshold $F$ in which
CEED has an abrupt change.\\
\begin{figure}
\includegraphics[width=8cm]{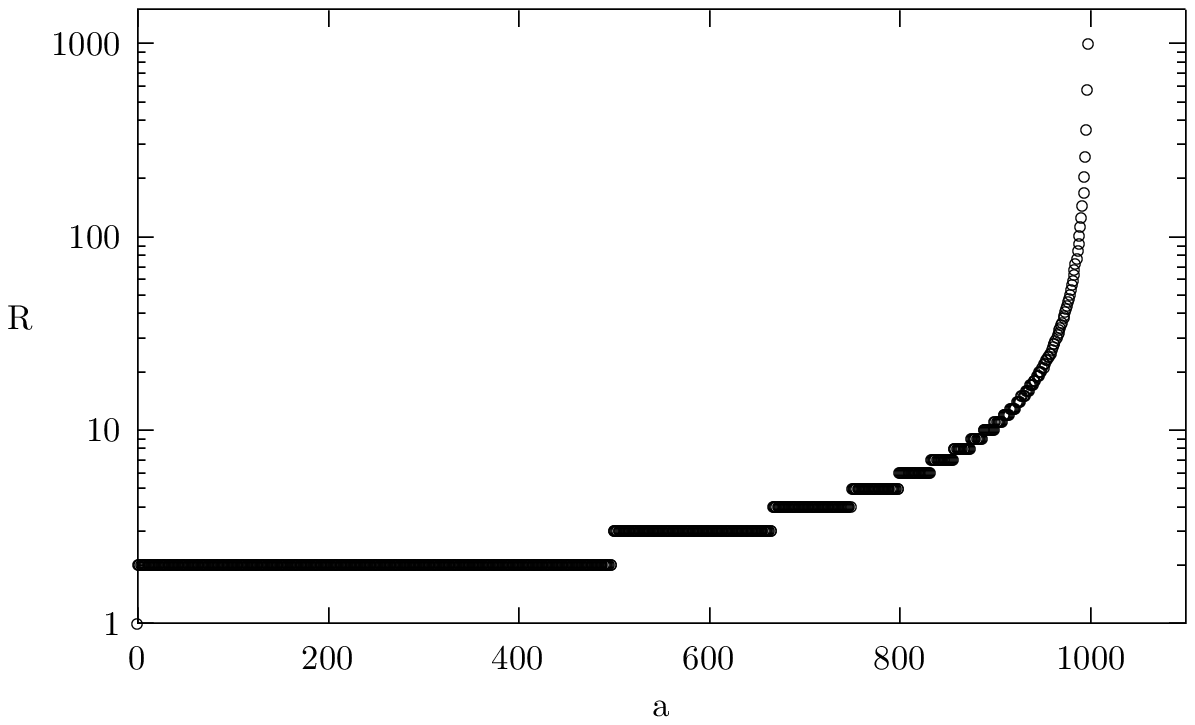}
\caption{ Chemical end-to-end distance vs $a$ for a complete
network of size $N=1000$ with $k=1$ and $k'=0$.}\label{f2}
\end{figure}
Next we consider the simple case of $k'=0$. It is clear that in
this situation, the shortcuts have no contribution in the elastic
properties of the network and we have a trivial problem in this
respect. But in the process of increasing $F$, the effect of
shortcuts in quantities like CEED is still important and
nontrivial. It is not difficult to show that in this limit
$P(\Delta_x) \propto (FN/k-\Delta_x)$. Moreover, as a function of
the number of avalanches $a$, we have
\begin{eqnarray}\label{cr}
F_a=\frac{k}{N-a}, \\ \nonumber R_a=1+[\frac{N}{N-a}]_+,\\
\nonumber M_a=M_0-\frac{a(a+1)}{2}, \\ \nonumber
K_a=\frac{k}{N-1},
\end{eqnarray}
where $[x]_+$ denotes the smallest integer larger than or equal to
$x$. In Fig. \ref{f2} we have shown $R$ versus $a$. Here again the
relation between force and CEED is linear. If we take $a$ and
$R_a$ as continuous variables and use Eq. (\ref{cr}) we obtain
\begin{equation}
P(\Delta R) \sim (\Delta R)^{-3/2}.
\end{equation}
This behavior is also seen in mean field models of fiber
bundles for the size distribution of avalanches \cite{s}.\\
For $0< k'<k$, $P(\Delta_x)$ still consists of three delta peaks.
Numerical simulations show that only for $k'\ll k$ the abrupt
change of CEED is replaced by a continuous one (more precisely a
staircaselike behavior). For example, if we choose $N=100$ and
$k=1$ we obtain a nearly continuous transition only for $k' <
0.01$. We found that this value of $k'$ is a decreasing function
of network size. Thus for $N \rightarrow \infty$ a
continues transition occurs only at $k'=0$.\\
\begin{figure}
\includegraphics[width=8cm]{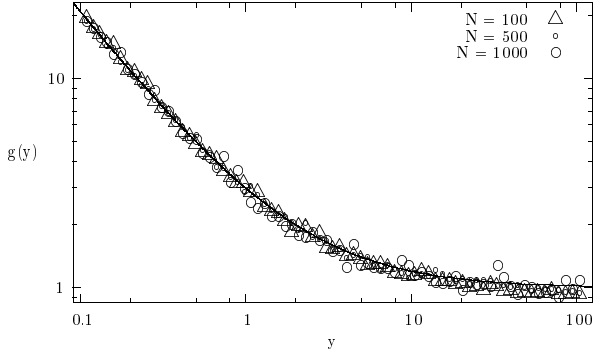}
\caption{ $g(y)=2K/(kpN)$ vs $y=pN^2$ for networks of size
$N=100,500$ (averaged over $1000$ realizations) and $N=1000$
(averaged over $300$ realizations). The line represents the curve
$1+2/y$.}\label{f3}
\end{figure}
\section{Random Small-world networks}\label{5}
We take a one-dimensional chain of $N$ nodes with $N-1$ regular
bonds. Then with probability $p$, we connect any two nodes of
distance larger than $1$ by a shortcut. In the following, we
consider three
different cases (i) $k=k'$, (ii) $k'=0$ and (iii) $0<k'<k$.\\
\subsection{The case $k=k'$}\label{51}
\begin{figure}
\includegraphics[width=8cm]{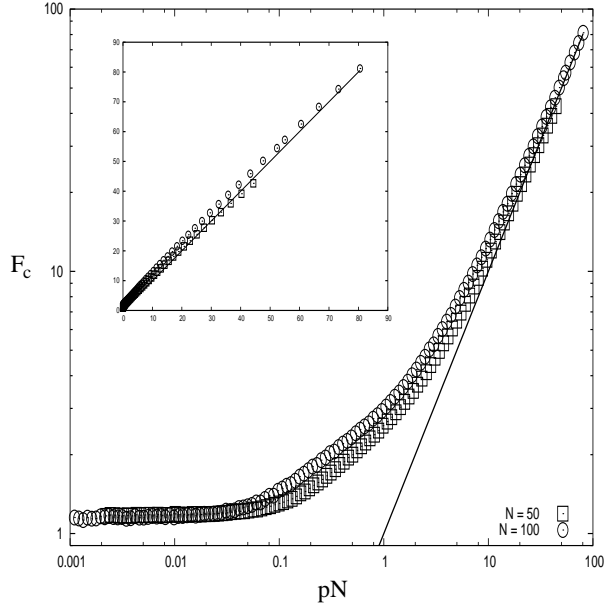}
\caption{ $F_c$ vs $pN$ in RSWNs with $k=k'=1$ for $N=50$ and
$N=100$ averaged over $5000$ and $1000$ realizations respectively.
The solid line displays $F_c=kpN$. In each step of the quasistatic
process we increase $F$ by $0.1$. The inset shows more clearly the
linear behavior of $F_c$ for large $pN$. }\label{f6}
\end{figure}

For $p> 1/N$ we expect to have $K=kpN/2$ which in the limit of $p=1$
is equal to the stiffness of a complete network. In other word if
$p> 1/N$,  we have effectively a complete network in which $k$ has been
replaced with $kp$. On the other hand for $p=0$ and large $N$
we have $K=k/N$. Using these limiting cases we suggest that
\begin{equation}
\frac{K}{N}=\frac{kp}{2}g(pN^2), \hskip 1cm g(y) \sim \left\{%
\begin{array}{ll}
    1, & \hbox{$y \gg 1$} \\
    \frac{2}{y}, & \hbox{$y \rightarrow 0$} \\
\end{array}%
\right.
\end{equation}
\begin{figure}
\includegraphics[width=8cm]{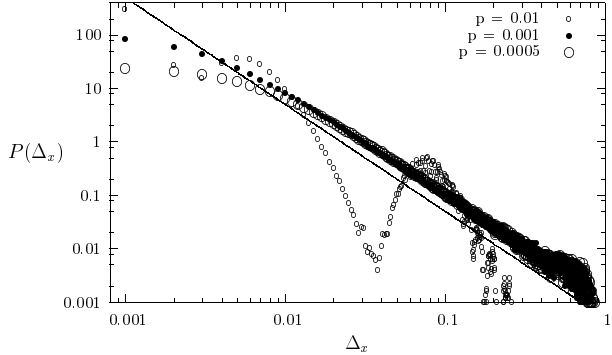}
\caption{Extension distribution for RSWNs of size $N=1000$ and
$k=k'=1$ averaged over $5000$ realizations. The line displays a
power law of exponent $-2$. }\label{f4}
\end{figure}
Note that for large $N$, $pN^2/2$ is equal to the average number
of shortcuts.
Numerical simulations shown in Fig. \ref{f3} support this scaling relation.\\
As $p$ decreases, we observe a crossover in $P(\Delta_x)$ from a three-modal behavior
to a scale free one, see Fig. \ref{f4}.
When $p> 1/N$ this distribution has three broad maximums instead of three
delta peaks in a complete network. The broadening of these peaks
is due to the random structure of the SWN. Just in the
small-world regime, $P(\Delta_x)$ is given by a power law
distribution of power $-2$. In this respect the RSWN belongs to the
universality class of the hierarchical model introduced in section
\ref{3}.\\
\begin{figure}
\includegraphics[width=8cm]{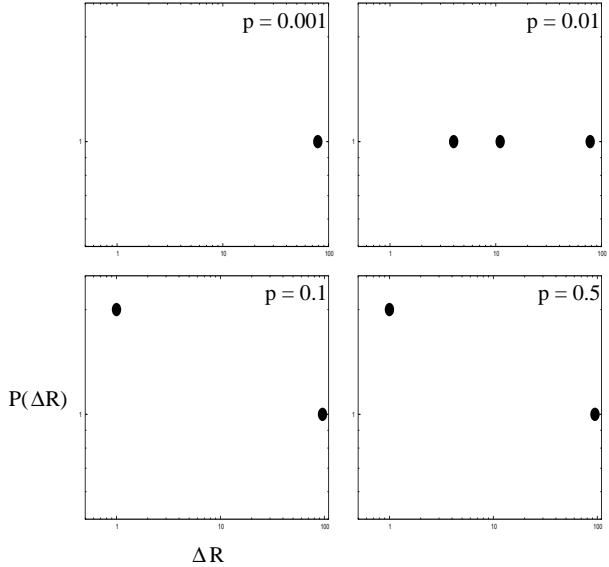}
\caption{ $P(\Delta R)$ for single realizations with different $p$
when $k=k'=1$, $N=100$. There are a few small changes in $R$
followed by a large change of order $N$ happened at critical force
$F_c$.}\label{f5}
\end{figure}
In our quasistatic process of increasing $F$, CEED changes nearly
abruptly at $F_c$. This behavior has been shown in Fig. \ref{f5}
that displays $P(\Delta R)$ for single realizations of the
process. We observe that after a few events of small size, we have
a large change of order $N$ in $R$ which we interpret it as a
discontinuous behavior of CEED. Again for $p\gg 1/N$,
$F_c=kp(N-2)$, see Fig. \ref{f6}. For $p\rightarrow 0$, when there
is at least one shortcut in the network, $F_c \simeq k \delta_c$.
These limiting behaviors of $F_c$ with $p$ have been shown in Fig.
\ref{f6}. We find that average number of shortcuts emanating from
a node, $pN$, determines $F_c$. As far as the shortcuts do not
overlap with each other, $F_c$ dose not depend on their number.
Actually for a finite number of shortcuts, a typical configuration
has almost nonoverlapping shortcuts. In this situation we have
$F_c \simeq k\delta_c$. Thus we expect that $F_c$ changes
considerably only when $pN \sim 1$. Moreover, from the linear
feature of the system we also expect that $F_c$ should be
proportional to $\delta_c$ and $k$. Thus, we suggest
$F_c=k\delta_c h(pN)$, where $h(pN)$ is a dimensionless scaling
function. In Fig. \ref{f6} we show that this relation works well
for two small-size networks and within our numerical
errors.\\
\begin{figure}
\includegraphics[width=8cm]{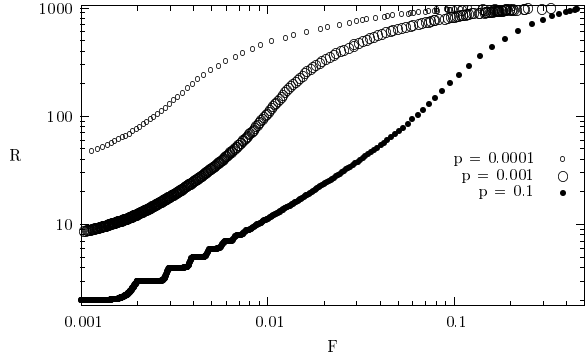}
\caption{The CEED vs $F$ in RSWNs of size $N=1000$, $k'=0$ and
$k=1$ after $5000$ (for $p=0.1,0.001$) and $10000$ (for
$p=0.0001$) realizations.}\label{f7}
\end{figure}
\subsection{The case $k'=0$}\label{52}
\begin{figure}
\includegraphics[width=8cm]{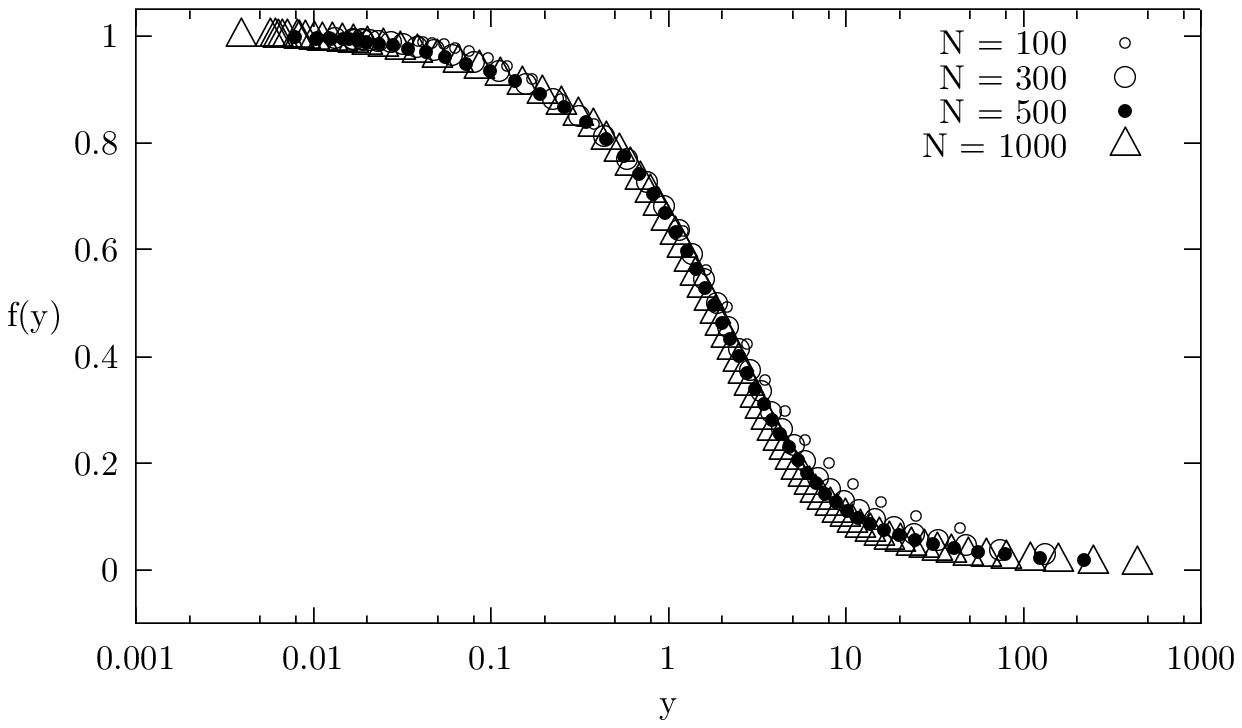}
\caption{$f(y)=R/N$ vs $y=pL_F^{2}$ for RSWNs of different sizes
(averaged over $10000$ realizations).}\label{f8}
\end{figure}

\begin{figure}
\includegraphics[width=8cm]{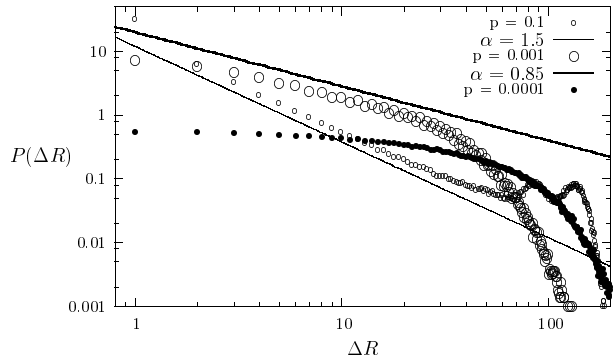}
\caption{ $P(\Delta R)$ for RSWNs of size $N=1000$, $k'=0$ and
$k=1$. The lines display $P(\Delta R) \propto (\Delta
R)^{-\alpha}$. Number of realizations are the same as those of
Fig. \ref{f6}.}\label{f9}
\end{figure}

When $k'=0$ the shortcuts have no contribution in the elastic
properties of the network. As indicated in the study of complete
networks, $P(\Delta_x)$ is a linear decreasing function. But
notice that during the quasistatic process, tearing of shortcuts
leads to considerable variations in $R$. In Fig. \ref{f7} we show
variation of $R$ versus $F$ obtained by numerical simulations.
There is no linear relation between $R$ and $F$. Indeed for a
given $F$, all shortcuts connecting two nodes of distance larger
than $L_F=k/F$ have been already teared. This introduces another
length scale. We expect $R$ to be a function of the number of
shortcuts in a subnetwork of size $L_F$, that is,
\begin{equation}
\frac{R}{N}=f(pL_F^2).
\end{equation}
Numerical simulations presented in Fig. \ref{f8} confirm this scaling relation.\\
Figure \ref{f9} displays $P(\Delta R)$
for various values of $p$. For a
small value of $p=0.1$, $P(\Delta R)\propto (\Delta R)^{-3/2}$
like a complete network. For smaller $p$,  we
find a power law distribution of lower exponent. For a very small
$p$, $P(\Delta R)$ is nearly a constant function with an
exponential tail.\\
\subsection{The case $0<k'<k$}\label{53}
\begin{figure}
\includegraphics[width=8cm]{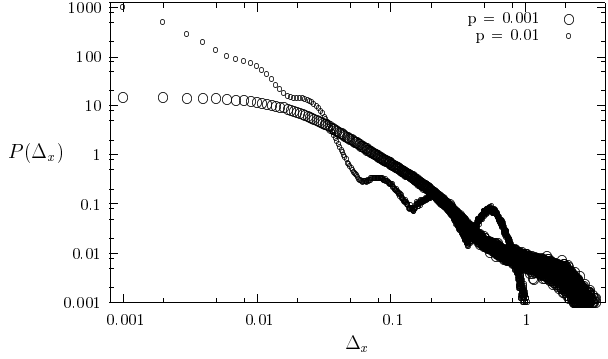}
\caption{Extension distribution for RSWNs of size $N=1000$, $k=1$
and $k'=0.1$ averaged over $5000$ realizations. }\label{f10}
\end{figure}

As before we take $k=1$ and decrease $k'$ from one. Numerical
simulations show that in this case the network stiffness is
proportional to $k'$. The effects of $k'$ and $p$ on elastic
properties are the same. It means that with respect to elastic
properties, the decreasing of $k'$ at a fixed $p$ is equivalent to
the decreasing of $p$ at a fixed $k'$. In Fig. \ref{f10} we
display the extension distribution when $k'=0.1$. It is observed
that when $p>1/N$ we have a multimodal distribution in contrast to
the three-modal distribution of case $k=k'$ for the same value of
$p$. Recall that we did not have such a behavior for lower values
of $p$ when $k=k'$. Indeed this new multimodal behavior
is observed only when we have a large number of shortcuts.\\
\begin{figure}
\includegraphics[width=8cm]{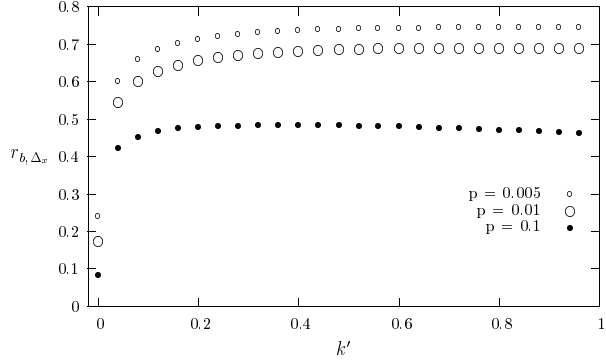}
\caption{Correlation coefficient of $\Delta x$ and $b$ for $N=100$
after averaging over $5000$ realizations.}\label{f11}
\end{figure}
The extension distribution of RSWNs shows that there are some shortcuts
bearing very large forces compared with the other ones. Certainly this is
because of their essential roles in the network structure. A good measure
of centrality of bonds in a network is their betweenness \cite{n1}. Suppose
that we have $n$ shortest pathes connecting node $1$ to $N$. A given shortcut may
contribute in $n_s$ of these pathes. Then, the betweenness of
this shortcut is defined as $b=n_s/n$. We define
\begin{equation}
r_{b, \Delta_x}:=\frac{<b \Delta_x
>-<b><\Delta_x>}{\sqrt{\sigma_b \sigma_{\Delta_x}}},
\end{equation}
as a measure of correlation between extension and betweenness of
shortcuts. In this definition $\sigma_{\Delta_x}$ and $\sigma_b$
are variances of $\Delta_x$ and $b$, respectively. In
Fig.\ref{f11} we show how the correlation coefficient depends on
$k'$ and $p$. As expected, $r_{b, \Delta_x}$ has a considerable
positive value in the small-world regime. The correlation
coefficient is nearly independent of $k'$ except for a rapid
decrease to zero
for $k' \rightarrow 0$.\\
\begin{figure}
\includegraphics[width=8cm]{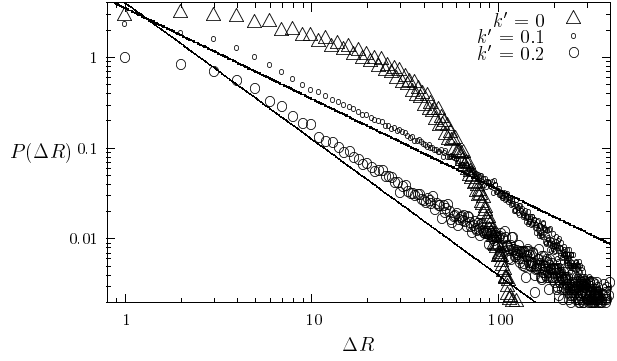}
\caption{ $P(\Delta R)$ for a network with $N=1000$, $p =0.001$
and $k=1$.  The data are results of averaging over $10000$
($k'=0$) and $5000$  ($k'=0.1,0.2$) realizations. The lines
represent  power laws of exponent $-1$ and $-3/2$.}\label{f12}
\end{figure}
\begin{figure}
\includegraphics[width=8cm]{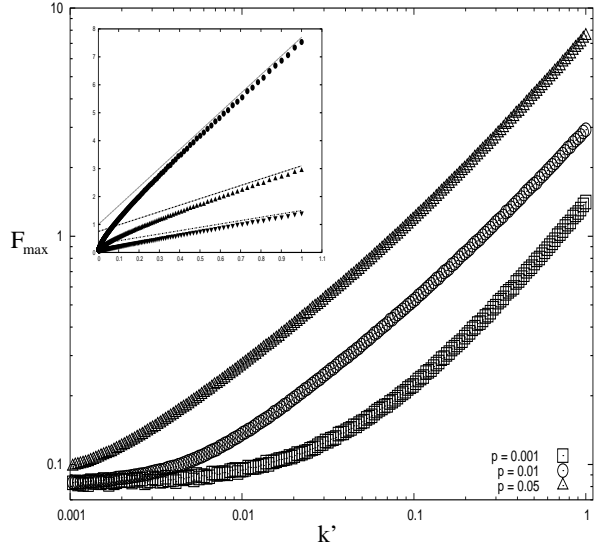}
\caption{ $F_{max}$ vs $k'$ for $N=100$ and $k=1$ (averaged over
$1000$ realizations). The inset shows more clearly the linear
behavior of data for large $k'$. The lines display linear
functions $F_{max}=a_0+a_1k'$. }\label{f13}
\end{figure}
In the quasistatic process, as it happens in complete networks, a
considerable number of avalanches takes place only for a
significantly small $k'$. Indeed, the larger $p$ the lower value
of $k'$; we need to see a continuous behavior of $R$. In Fig.
\ref{f12} we compare $P(\Delta R)$ for two cases of $k'=0$ and
$k'=0.1$. The figure shows that as expected, by increasing $k'$,
$P(\Delta R)$ approaches to that of a complete network. Indeed, in
the small-world regime and for a nonzero $k'$, $P(\Delta R)$ obeys
a power law of exponent $-\alpha$ with $\alpha < 3/2$. This
exponent
approaches to $-3/2$ as we increase $k'$ or $p$.\\
In this case it is not easy to define $k'_c(p)$, the value which
separates continuous and discontinuous regimes. Instead we
calculate $F_{max}$, the force which leads to $R_{max}=N-1$. We
expect that a change in the behavior of $F_{max}$ with respect to
$k'$ signals a crossover from a continuous to a discontinuous
region. In Fig. \ref{f13} we show the variation of $F_{max}$
versus $k'$ obtained by numerical simulation. We see that by
decreasing $k'$, the linear behavior of $F_{max}$ changes and then
it saturates for $k'\rightarrow 0$. This crossover occurs for a
lower $k'$ as one enhances the number of shortcuts.
\section{Conclusions}\label{6}
\begin{figure}
\includegraphics[width=8cm]{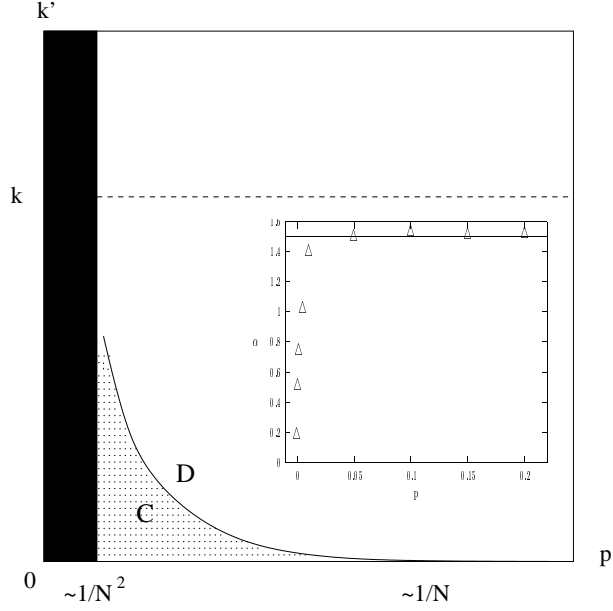}
\caption{ Schematic representation of phase diagram. Dashed region
indicates where we have a continuous behavior of $R$ vs. $F$. For
$p<1/N^2$ (black region) we can not speak of continuous or
discontinuous behavior of $R$. Note that as $N$ approaches to
infinity the continuous region reduces to the line $k'=0$. The
inset shows the variation of $\alpha$ with $p$ when $N=1000$,
$k=1$ and $k'=0$. The errors in $\alpha$ are of order $0.01$. The
same behavior will be observed for $\alpha$ when we fix $p$ and
increase $k'$. The solid line in inset indicates the expected
value of $\alpha$ for large $p$ i.e. $3/2$.}\label{f14}
\end{figure}

We studied the static and quasistatic properties of SWNs. We
observed that the network's structure significantly affects the
static and quasistatic behavior. Thus,
we could draw some conclusions about the structure of networks through the study
of these properties.\\
The summary of the main results for RSWNs have been represented in
table \ref{t} and Fig. \ref{f14}. We found that for very small
values of $p$, the effective stiffness of the network is
comparable with that of a complete network. When $k=k'$ the
extension distribution of shortcuts is a power law of exponent
$-2$. In this respect RSWNs behave like a hierarchical network
introduced in this paper. There was also a strong positive
correlation between the betweenness of a shortcut and its
extension. It means that just by looking at the distribution of
extensions in a network, one could be
able to distinguish which shortcuts are more central.\\
In the quasistatic part, we showed that by increasing $F$, CEED
could have a continuous or discontinuous transition. In general,
to have a continuous transition we need a much smaller spring
constant for shortcuts rather than regular bonds. In the case of a
discontinuous transition, the critical force is determined by the
average number of shortcuts per node. It was found that for a
continuous transition, $P(\Delta R)$ is given by power law
distributions
of the exponent $-\alpha$ with $\alpha \le 3/2$.\\
\acknowledgments We would like to thank S. N. Rasuli for useful
discussions. We are grateful to MirFaez Miri for comments and
careful reading of the manuscript.

\begin{table}

\begin{center}

\begin{tabular}{|c|c|c|}
  \hline

                 &  $P(\Delta_x)$ & $P(\Delta R)$   \\
  \hline
     $k=k'$      & $(\Delta_x)^{-2}$ & $\delta_{\Delta R,N}$ (an abrupt change of R) \\
  \hline
     $k'=0$      & $a_0-a_1\Delta_x$ & $(\Delta R)^{-\alpha} (\alpha \leq3/2)$   \\
  \hline
     $0<k'<k$    & $(\Delta_x)^{-\beta} (\beta \leq 2)+$ oscillations (for large $p$) & $\left\{%
\begin{array}{ll}
    (\Delta R)^{-\alpha} (\alpha \leq3/2), & \hbox{$k' < k'_c(p) $} \\
    \delta_{\Delta R,N}, & \hbox{ $k' > k'_c(p) $} \\
\end{array}%
\right\} $  \\

  \hline

\end{tabular}

\vskip 0.5cm

\caption{Summary of results for $P(\Delta_x)$ and $P(\Delta R)$ in
different cases. Here $k'_c(p)$ gives the line in $(p,k')$ plane
that separates continuous and discontinues behavior of
CEED.}\label{t}
\end{center}
\end{table}

\end{document}